\documentclass[preprint,aps]{revtex4}
\usepackage{graphicx}
\usepackage{dcolumn}
\usepackage{bm}

\begin{document}

\title{Controlled Dephasing of a Quantum Dot:\\ From Coherent to Sequential Tunneling}

\author{Daniel Rohrlich$^1$}
\altaffiliation[Current address: ]{Department of Physics, Ben-Gurion University, Beersheva
84105 Israel.}
\author {Oren Zarchin$^1$}
\altaffiliation[E-mail: ]{{\tt oren.zarchin@weizmann.ac.il}
\\${^1}$Equal contribution }
\author {Moty Heiblum}
\author {Diana Mahalu}
\author {Vladimir Umansky}
\address{Braun Center for Submicron Research, Department of Condensed
Matter Physics,\\ Weizmann Institute of Science, Rehovot 76100
Israel}

\date{\today}

\begin{abstract}
     Resonant tunneling through identical potential barriers is a textbook problem in
quantum mechanics.  Its solution yields total transparency (100\% tunneling) at discrete
energies. This dramatic phenomenon results from coherent interference among many
trajectories, and it is the basis of transport through periodic structures.  Resonant
tunneling of electrons is commonly seen in semiconducting ``quantum dots".  Here we
demonstrate that detecting (distinguishing) electron trajectories in a quantum dot (QD)
renders the QD nearly insulating. We couple trajectories in the QD to a ``detector" by
employing edge channels in the integer quantum Hall regime.  That is, we couple electrons
tunneling through an inner channel to electrons in the neighboring outer, ``detector"
channel.  A small bias applied to the detector channel suffices to dephase (quench) the
resonant tunneling completely.  We derive a formula for dephasing that agrees well with
our data and implies that just a few electrons passing through the detector channel
suffice to dephase the QD completely. This basic experiment shows how path detection in a
QD induces a transition from delocalization (due to coherent tunneling) to localization
(sequential tunneling).
\end{abstract}


\maketitle

     The study of entanglement began in 1935 with the EPR \cite{epr} and Schr\"odinger Cat
\cite{scat} paradoxes, but it languished until Bell's celebrated 1964 paper \cite{bell}
and even thereafter.  More recently, applications of entanglement to cryptography
\cite{qcode}, ``teleportation" \cite{tp}, data compression \cite{qdc} and computation
\cite{qcom} have given new impetus to the study of entanglement.  Also the {\it loss} of
interference (``decoherence" or ``dephasing") is studied, both as a condition for
classical behavior to emerge from quantum systems and, more recently, as an obstacle to
applications of entanglement.  Here we report controlled partial and full dephasing of
electron interference in a mesoscopic Fabry-Perot type interferometer---a quantum dot
(QD)---entangled efficiently to a mesoscopic detector.

     Mesoscopic interferometers \cite{icqox} include closed \cite{yacoby} and open
\cite{schus} two-path interferometers, QDs and double-QDs
\cite{sprinzak}, and electronic Mach-Zehnder interferometers
\cite{ji}.  Mesoscopic detectors \cite{icqox} include quantum point
contacts (QPCs) \cite{fields,buks} and partitioned currents
\cite{sprinzak}. In our experiment, a QD serves as an interferometer
of the Fabry-Perot type; the interference shows up as a resonant
transmission peak in electron conductance through the dot.
\hbox{Figure 1} shows the QD.  In order to couple tunneling and
detector electrons strongly, we chose them from neighboring edge
channels (i.e. in close proximity) in the integer quantum Hall
regime. We worked at filling factors $\nu$=2 and $\nu$=3, but
nothing in our results depends essentially on edge channels or a
magnetic field. For the innermost quantum Hall edge channel (i.e.
the channel farthest from the boundary) the dot is an
interferometer. As electrons in the innermost channel tunnel through
the dot, they become entangled with electrons passing freely through
the neighboring, outer edge channel, which serves as a ``detector"
channel. These detector electrons couple coulombically to the total
charge $Q_{tun}$ tunneling through the dot, and their accumulated
phase is proportional (via this Coulomb coupling) to the dwell time
$t_{dwell}$ of the tunneling electrons: $Q_{tun} =t_{dwell}
I_{tun}$, where $I_{tun}$ is the tunneling current. Detection
broadens and quenches the resonance, consistent with the time-energy
uncertainty principle:  the decreased uncertainty in the dwell time
entails increased uncertainty in the energy of the electrons.

     According to a general principle \cite{feyn}, any determination of the path an electron
takes through an interferometer, among all possible interfering paths, destroys the
interference among the paths. Hence, coupling (entangling) a trajectory-sensitive detector
and an electron interferometer should destroy the interference.  In our experiment the
detector is a partitioned channel current; it is partitioned at a quantum point contact
(QPC) (not shown in Fig. 1) before reaching the QD.  Why partitioned?  The detector
current acquires a phase due to Coulomb coupling with the tunneling electrons in the inner
channel. However, if the detector current is full (unpartitioned, noiseless) this phase is
unobservable. Partitioning the detector current produces a transmitted {\it and} a
reflected current; these two currents could interfere elsewhere and render the
unobservable phase observable. Hence, partitioning the detector current allows us, in
principle, to extract the additional phase due to coupling with electrons tunneling
through the quantum dot. Now, whether or not we actually interfere the transmitted and
reflected currents {\it elsewhere} cannot instantly produce any measurable change at the
dot. Hence, a partitioned current must by itself dephase the electron resonance in the
interferometer.

     In this account, dephasing arises because the interfering quanta (electrons in the dot)
leave ``which path" information in the environment (detector current).  Yet according to
another general principle \cite{sai}, there is always a complementary account: dephasing
arises because the environment (detector current) produces fluctuating phases in the
interfering quanta, and thus dephases the resonance. The partitioned current fluctuates:
if $N$ electrons arrive at a QPC that transmits with probability $T$, then $NT$ are
transmitted, on average, with typical fluctuations of $\sqrt{NT(1-T)}$. These fluctuations
in the detector current (``shot noise" \cite{noise}) produce a fluctuating potential at
the dot and thus a fluctuating phase in the tunneling electrons, which dephases the
resonance.

     For a Fabry-Perot interferometer, we can model the dephasing by calculating the effect
of detection on interference.  Let the first and second QPCs of the dot transmit with
amplitudes $t_1$ and $t_2$ and reflect with amplitudes $r_1$ and $r_2$, respectively.  In
the absence of a fluctuating phase, the amplitude $t_{tun}$ for resonant transmission
through the dot would be
\begin{equation}
t_{tun}= t_1 t_2 \left[ e^{i\theta} +(r_1 r_2 )e^{3i\theta} +(r_1 r_2 )^2 e^{5i\theta}
+\dots \right] = t_1t_2 \sum_{j=0}^\infty (r_1 r_2 )^j e^{i(2j+1)\theta}~~~;
\label{series}
\end{equation}
the sum includes an energy-dependent phase $2\theta$ for each back-and-forth lap in the
interferometer.  However, we assume that during each back-and-forth lap, $N$ electrons
reach the QPC that partitions the detector current. Each transmitted detector electron
induces an additional phase $\epsilon$ to a single back and forth trajectory of the
resonant tunneling electron, while reflected detector electrons do not affect the
tunneling electron.  Indexing the detector electrons $k=0, 1, 2, \dots$ according to their
order of arrival at the detector QPC, we have additional phases $\epsilon_k$ where
$\epsilon_k =\epsilon$ if the $k$-th electron is transmitted through the QPC and
$\epsilon_k =0$ if it is reflected. Then for a given partitioning of the detector current
we obtain not Eq.\ (\ref{series}) but
\begin{equation}
t_{tun}= t_1t_2 \sum_{j=0}^\infty (r_1 r_2 )^j e^{i(2j+1)\theta}
e^{i (\epsilon_0 +\epsilon_1+\dots +\epsilon_{j\cdot N})} ~~~~.
\label{revised}
\end{equation}
Actually, Eq.\ (\ref{revised}) lacks the phase due to the first $N/2$ detector electrons
to reach the detector QPC (i.e. as the tunneling electron first crosses the
interferometer), but since this phase is common to all the terms in the sum, we neglect
it. The transmission probability, given this partitioning, is the square of the absolute
value of Eq.\ (\ref{revised}):
\begin{equation}
T_{tun} =\vert t_{tun}\vert^2  = T_1T_2 \sum_{j, j^\prime =0}^\infty (r_1 r_2 )^j (r^*_1
r^*_2 )^{j^\prime} e^{2(j-j^\prime )i\theta} e^{i \sum_{k=0}^{j\cdot N} \epsilon_k -i
\sum_{k^\prime =0}^{j^\prime \cdot N} \epsilon_{k^\prime} }~~~, \label{thissum}
\end{equation}
where $T_1 =\vert t_1\vert^2$, etc.  We have to fold Eq.\ (\ref{thissum}) with the
probability distribution for the given partitioning of detector electrons.  We do so in
two steps. First, for a fixed $j-j^\prime \ge 0$ in Eq.\ (\ref{thissum}), we sum over
$j^\prime$; that is, we consider
\begin{equation}
T_1T_2 \sum_{j^\prime =0}^\infty (r_1 r_2 )^{j-j^\prime} (R_1 R_2
)^{j^\prime} e^{2(j-j^\prime )i\theta} e^{i \sum_{k=j^\prime \cdot
N}^{j\cdot N} \epsilon_k}~~~~. \label{partsum}
\end{equation}
We now fold the distribution of phases $\epsilon_k$ into Eq.\ (\ref{partsum}) by replacing
$e^{i \sum_{k=j^\prime \cdot N}^{j\cdot N} \epsilon_k}$ with \hbox{$(e^{i\cdot 0}R+e^{i
\epsilon} T)^{(j-j^\prime)N}$,} where $R$ and $T$ are, respectively, the probability for
reflection and transmission of electrons from the detector QPC \cite{eps}. After summing
over $j^\prime$ in Eq.\ (\ref{partsum}), the next step is to sum over all values of
$j-j^\prime$.  (Note that for $j-j^\prime\le 0$, we replace Eq.\ (\ref{partsum}) by its
complex conjugate.)  The resulting transmission probability, which we denote $\langle
T_{tun} \rangle$ to indicate the averaging over detector partitionings, is
\begin{equation}
\langle T_{tun} \rangle ={{T_1T_2}\over{1-R_1R_2}} \left[ {1\over
{1-M}} + {1\over {1-M^*}} -1\right] = {{T_1T_2}\over{1-R_1R_2}}
{{1-M^* M}\over {\vert 1-M\vert^2}}~~~, \label{formula}
\end{equation}
where $M\equiv e^{2i\theta} r_1r_2 (R+e^{i\epsilon}T)^N $.  The integral of $\langle
T_{tun} \rangle$ over ${\bar\theta} \le \theta \le ({\bar\theta} + \pi )$, for any real
$\bar\theta$, is independent of $\vert M\vert$ (as it must be since probabilities must sum
to 1 for any strength of dephasing).  For $R_{1,2} \gg T_{1,2}$ and small $\epsilon$, Eq.\
(\ref{formula}) implies both broadening and quenching (decreased height) of the resonance
peak in proportion to $NT(1-T)$, as derived before \cite{davies}. Here, however, with the
detector and tunneling currents so close, we cannot assume $\epsilon$ small.

     The device, constructed from a GaAs-AlGaAs heterojunction (see Fig. 1), supported a
high-mobility two-dimensional electron gas (2DEG).  Biased metallic gates deposited on the
surface of the heterojunction induced a controlled backscattering potential to form the
quantum dot and quantum point contacts.  The magnetic field was 5-7 Tesla, well within the
filling-factor 2 conductance plateau.  Conductance was measured with a 0.9 MHz AC, 0.5
$\mu$V rms excitation voltage at an electron temperature of $\tau =$ 25 mK. A low-noise
cryogenic preamplifier in the vicinity of the sample amplified the measured voltage,
followed by a room-temperature amplifier and a spectrum analyzer. An {\it LC} resonant
circuit prior to the cold preamplifier allowed measurement of the signal at about 0.9 MHz
with a bandwidth of about 100 Hz; see Ref. \cite{picc} for details.

     Figure 2 shows dephasing of a series of Coulomb blockade peaks for various
partitionings $T$ of the detector current, at detector bias $\it{V_D}=77 \mu$V. For the
horizontal axes we convert plunger gate potential to an effective dot potential (a
``levering factor" extracted from Coulomb-diamond measurements \cite{lever}). The
resonance peaks quench and broaden as $T$ increases from 0 to 1/2 and reemerge as $T$
increases from 1/2 \hbox{to 1.} At $T=0$ there is no current in the detector to dephase
the resonance.  At $T=1$ the resonance induces a constant phase in the electrons of the
detector current, but the phase is not observable and there is again no dephasing.  Only
when $T$ is between these limits does the detector current contain information about the
resonance, and dephases it. Smaller detector bias implies less information in the detector
current (or, in the complementary account, less shot noise in the detector current) hence
less dephasing. Indeed, resonance peaks are less quenched at smaller detector bias.

     Looking in detail at one conductance peak and fitting it with a Lorentzian curve, we
obtain the width of the resonance peaks (Fig 2b). Undephased peaks have a full width at
half maximum (FWHM) of about 12 $\mu$V, larger than $4k_B \tau \approx 9~\mu$V (where
$k_B$ denotes the Boltzmann constant) at an estimated electron temperature of $\tau =$ 25
mK. We found that $T$ depended slightly on the detector bias.  Thus, for each value of
detector bias, we have calculated an effective transmission $T_{eff}$ by averaging $T$
with respect to energy, from the Fermi energy to the detector bias, and Fig. 3 shows
dependence of (a) peak height and \hbox{(b) peak} width on $T_{eff}$, with the bias on the
detector as an additional parameter.

     To understand the relation between shot noise and dephasing quantitatively, let us
define three times: $t_{dwell}$, $t_{lap}$ and $t_{det}$.  In the absence of temperature
broadening, the dwell time $t_{dwell}$ would be $\hbar$ divided by $12 ~\mu$V, the FWHM of
the resonance. However, the FWHM is a convolution of coherent broadening and temperature
broadening; only the former is relevant to the dwell time. Subtracting the temperature
broadening $4k_B \tau \approx 9~\mu$V from $12 ~\mu$V we are left with $3 ~\mu$V, so
$t_{dwell}\approx \hbar /3\mu V \approx $220 psec.

     The dwell time is a multiple of the lap time, i.e. the time $t_{lap}$ it takes an
electron to go once back and forth in the dot. How many laps in a dwell time?  To answer
this question we return to Eq.\ (\ref{series}) and note that a term $t_1t_2 (r_1 r_2)^j
e^{i(2j+1)\theta}$ in the series corresponds to $j+1/2$ laps.  Then the average number of
laps made by an electron tunneling through the resonance is $\sum_j (j+1/2) T_1T_2
(R_1R_2)^j$ divided by the total probability $\sum_j T_1T_2 (R_1R_2)^j$ to tunnel through
the resonance, so it equals $1/2 + R_1R_2/ (1-R_1R_2)$.  In our experiment, we estimate
$R_1\approx R_2 \approx 2/3$ and so the average number of laps was approximately 1.3, i.e.
the most likely path of an electron tunneling through the dot was to reflect twice inside
the dot.  Dividing $t_{dwell}$ by the average number of laps, we obtain $t_{lap} \approx
170$ psec as the lap time.  (From $t_{lap}$ we can estimate the speed of an electron
tunneling through the dot: if the effective inner length of the dot was roughly 0.25
$\mu$m, then the electron traveled 0.5 $\mu$m in 170 psec, i.e. its speed was roughly
$3\cdot10^5$ cm/sec, corresponding to a rather small electric field in the dot.)

     Finally, the time $t_{det}$ between successive electrons in the unpartitioned detector
current $I$ is $e/I=eR_H/V$ where $R_H$ is the Hall resistance $R_H=h/e^2$ and $V$ is the
bias applied to the detector.  Thus, $t_{det} =h/eV$, which was as low as 40 psec for the
maximum detector bias of 103 $\mu$V.  For this maximum detector bias, an average of $N$=
170 psec/40 psec electrons, i.e between 3 and 4 detector electrons reached the detector
QPC during each lap of the tunneling electron; the number is proportionally smaller for
smaller bias. This number corresponds to $N$ above in Eqs.\ (\ref{revised}-\ref{formula}).
Taking $N$ to be proportional to the detector bias potential $V$, we find experimentally
that the broadening and quenching of the resonance peak are both proportional to the shot
noise $NT_{eff} (1-T_{eff})$ at low detector bias (10 $\mu$V $\le$ V $\le$ 50 $\mu$V) but
deviate from simple proportionality at larger bias (Fig. 3a).  In particular, at larger
bias, quenching of the resonance peak tends to saturate before $T_{eff}$ reaches 0.5. This
saturation is just what Eq.\ (\ref{formula}) implies, since the peak height (i.e. the
difference between maximum and minimum values of $\langle T_{tun} \rangle$ as a function
of $\theta$) obtained from Eq.\ (\ref{formula}) is
\begin{equation}
{{T_1T_2}\over{1-R_1R_2}}   \left[ {{1-Z}\over{\left(
1-\sqrt{Z}\right)^2}} - {{1-Z}\over{\left( 1+\sqrt{Z}\right)^2}}
\right] ={{T_1T_2}\over{1-R_1R_2}}{4\sqrt{Z}\over{1-Z}}~~~,
\label{ph}
\end{equation}
where $Z\equiv R_1 R_2[1+2RT(\cos \epsilon -1)]^N$.  For small $\epsilon$, Eq.\ (\ref{ph})
reduces to
\begin{equation}
{{4T_1T_2\sqrt{R_1 R_2}}\over{(1-R_1R_2)^2}} \left[ 1-\left({1\over 2} + {{R_1R_2}\over
{1-R_1R_2}}\right) NRT\epsilon^2 +{\cal O} (\epsilon^4) \right]~~~, \label{phe}
\end{equation}
so for small $\epsilon$ the peak height depends linearly on shot noise $NRT=NT(1-T)$ and
quadratically on $T$, as noted above \cite{laps}.  But when $\epsilon$ is not small, Eq.\
(\ref{ph}) tends to saturate in $T$ for large bias (large $N$), as the fits to Fig. 3(a)
show.

     Equation\ (\ref{formula}) also leads to a formula for the broadening of the
resonance peak as a function of detector bias and partitioning:
\begin{equation}
{\rm FWHM} = {\hbar \over t_{lap}} \arctan {1\over 2}\left[ {1\over {\sqrt{Z}}} - \sqrt{Z}
\right] ~~~~. \label{fwhm}
\end{equation}
Equation\ (\ref{fwhm}) implies saturation of broadening before $T_{eff}(1-T_{eff})$
reaches its maximum value, for large bias. Yet Fig. 3(b) indicates ``anti-saturation" in
$T_{eff}$: that is, the data do not level off in the middle of the range of $T_{eff}$ but
cluster upwards in the form of a triangle.  This apparent inconsistency with our model may
be understood as an artifact of the multiplicity of peaks.  Each peak is enhanced by the
tails of its neighbors, and this enhancement increases with the increased dephasing of the
peaks.  The enhancement does not significantly affect the apparent height of a peak, which
is measured farthest from the neighboring peaks, but strongly affects apparent broadening.
In addition, a Fabry-Perot resonance is equivalent to a Lorentzian only near the peak.
Hence we have not applied \hbox{Eq.\ (\ref{fwhm})} to Fig. 3(b) for the largest bias.

     Additional support for our analysis of dephasing comes from measurements which we made
on the same mesoscopic device, but with another setup at filling factor 3 and electron
temperature of $\sim100$ mK.  These measurements checked the dependence of dephasing on
the magnetic field at $B=4.0$ T and $B=4.3$ T, within the $3e^2/h$ conductance plateau.
Since the $\nu =2$ and $\nu =3$ edge channels are separated by a cyclotron gap, we expect
large channel separation and weaker dephasing, in accord with the small-$\epsilon$ limit
of Eq.\ (\ref{formula}).  For small $\epsilon$, Eq.\ (\ref{fwhm}) implies a broadening in
FWHM proportional to $NRT\epsilon^2$. Indeed, we found that the FWHM depended linearly on
$I T_{eff} (1-T_{eff})$ and that the slope of the line was some 40\% higher at $B=4.0$ T
than at $B=4.3$ T.

     In summary, we have demonstrated controlled dephasing of a resonant tunneling device
(a quantum dot) and showed how the dephasing depends on the detector current and
partitioning.  Controlled dephasing was realized in the integer quantum Hall regime, where
we exploited the close proximity of edge channels to strongly entangle a small number of
electrons.

\bigskip

\acknowledgments We thank Yang Ji, Yunchul Chung, Michal Avinun and Izhar Neder for
technical help and Florian Marquardt and Izhar Neder for helpful discussions.  O. Zarchin
acknowledges support from the Israeli Ministry of Science and Technology.  This work was
partly supported by the MINERVA foundation, the German Israeli foundation (GIF), the
German Israeli project cooperation (DIP), and the Israeli Science foundation (ISF).

\begin{figure}[H!]
\begin{center}
        \label{fig1}
        \includegraphics[width=5in]{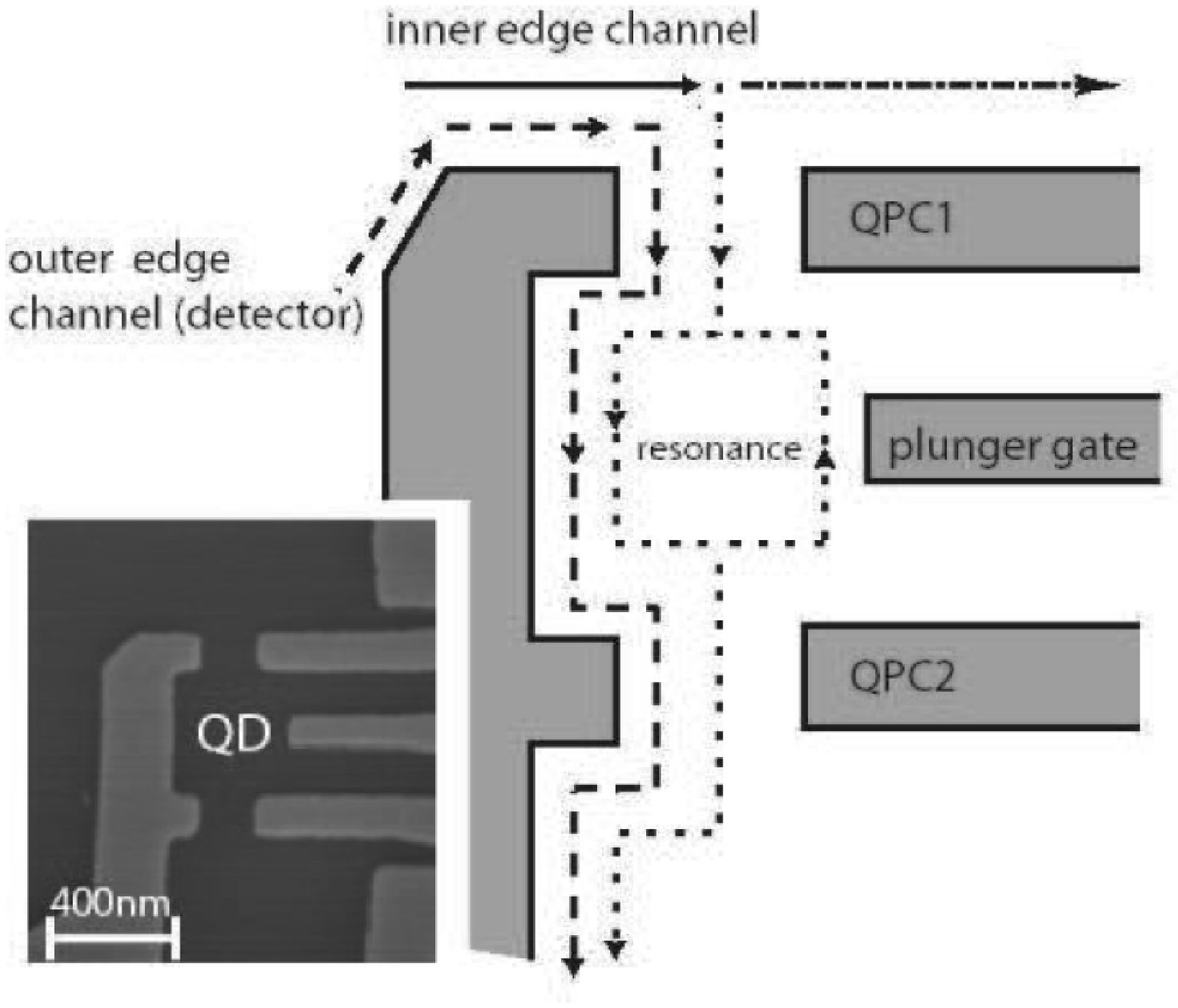}
\end{center}
\caption{Diagram of the quantum dot, defined by biased metallic
electrodes (two QPCs and a ``plunger gate") over a high-mobility
two-dimensional electron gas (2DEG) of density 2 $\times 10^{11}
/$cm$^2$ embedded in a GaAs-AlGaAs heterojunction.  At magnetic
field 5-7 T the 2DEG is at the filling-factor 2 plateau. Two quantum
Hall edge channels enter from above.  The inner current channel
crosses via resonant tunneling and the outer current, partitioned at
a prior quantum point contact, serves as a detector.  Inset:  SEM
micrograph of a similar dot, 0.4 $\mu m$ wide inside.}
\end{figure}

\begin{figure}[H!]
\begin{center}
        \label{fig2}
        \includegraphics[width=5in]{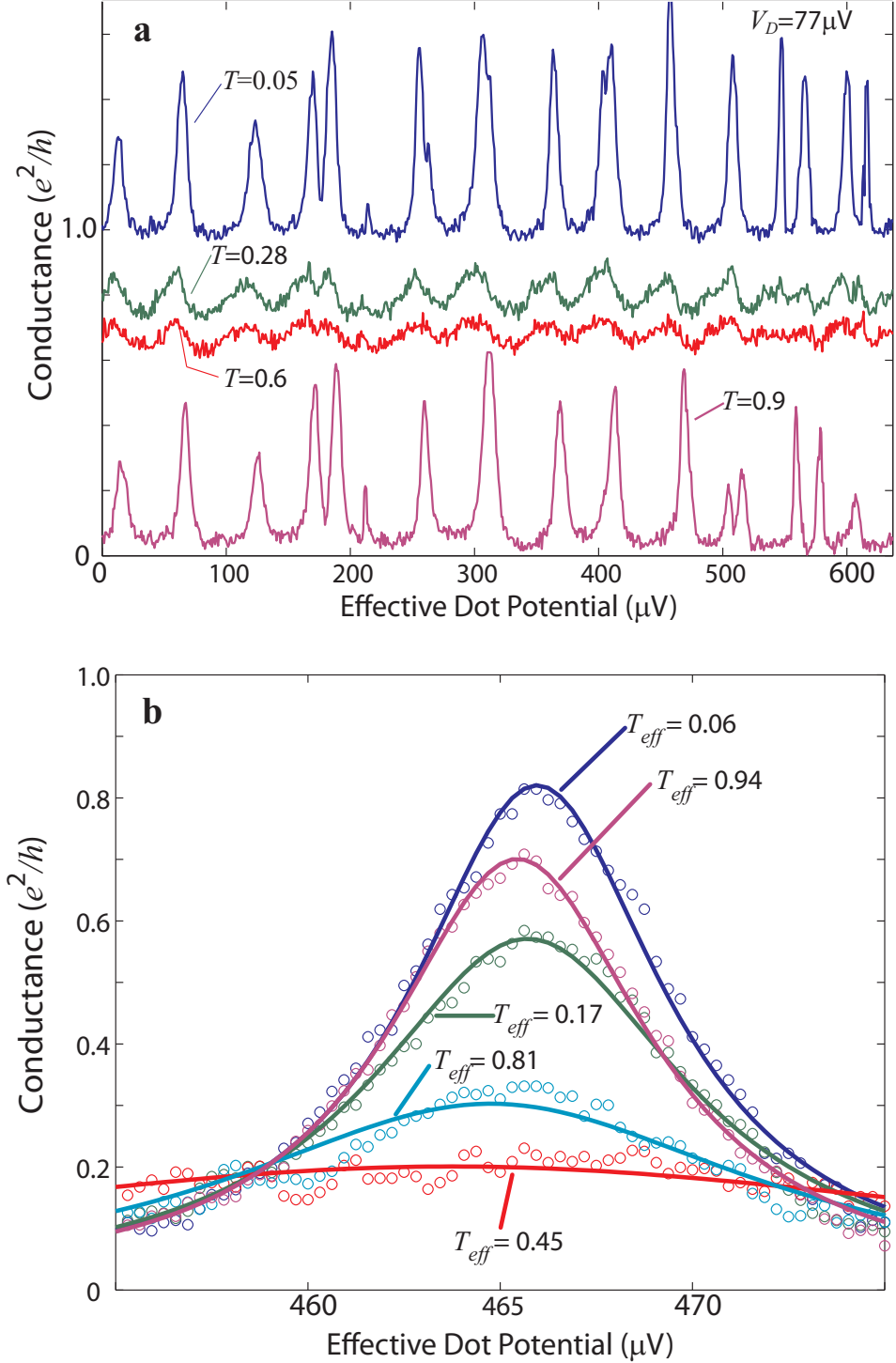}
\end{center}
\caption{Dephasing of resonance peaks at filling factor 2, with 77
$\mu$V DC bias on the detector.  \hbox{(a) The} horizontal axis
shows the potential on the plunger gate, normalized to effective dot
potential.  The vertical axis shows the resonant conductance through
the inner channel (shifted), ranging from $T \approx 0$ (top trace)
to $T \approx 1$ (bottom trace).  (b) Dephasing of a typical
resonance peak. The vertical axis shows the resonant conductance.
Circles are experimental results while lines are Lorentzian fits.}
\end{figure}

\begin{figure}[!h]
\begin{center}
        \label{Fig3}
        \includegraphics[width=5in]{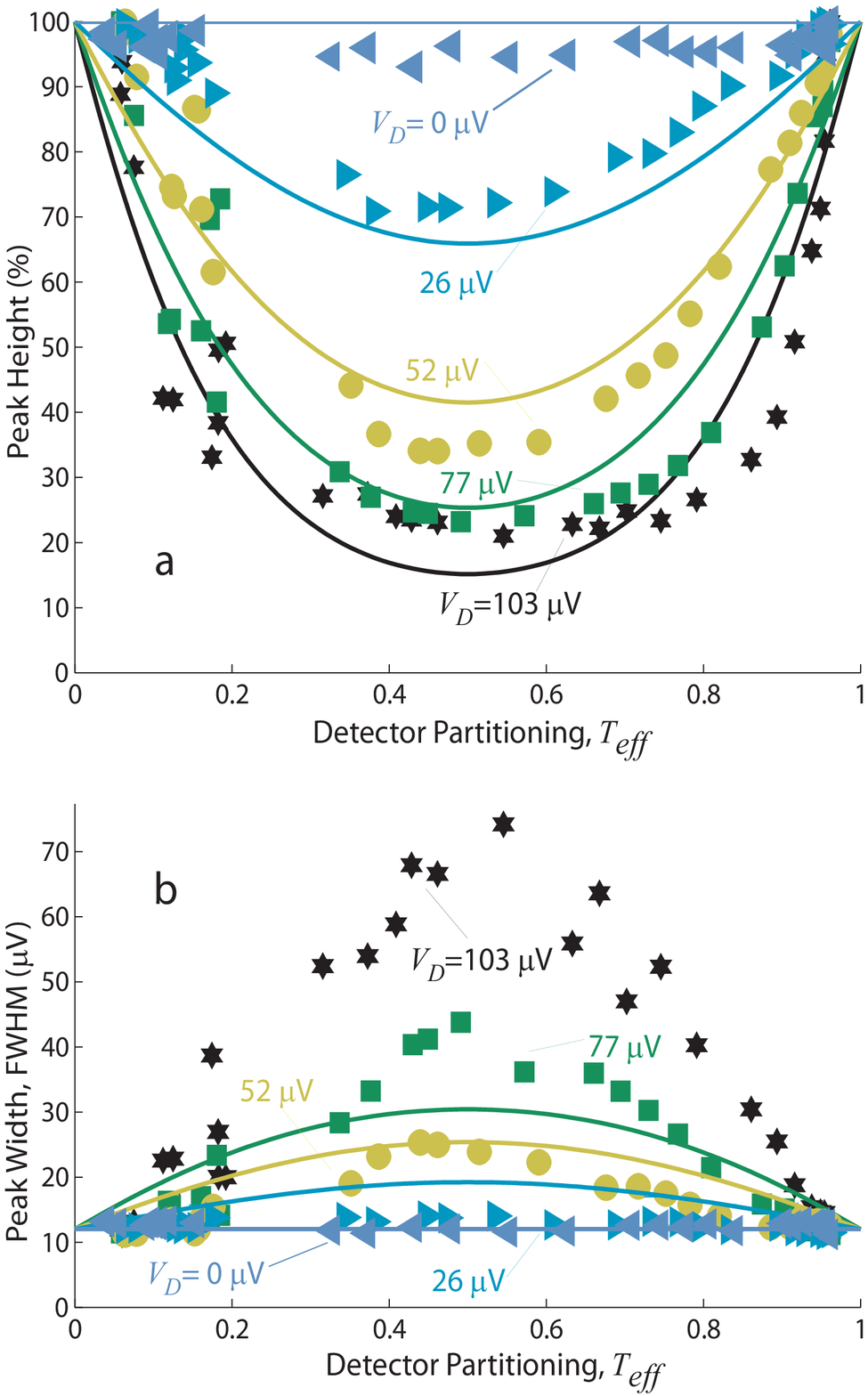}
\end{center}
\caption{Plots of (a) quenching (as percentage of the original peak
height) and (b) peak width FWHM, as functions of the effective
detector transmission $T_{eff}$ and detector bias.  Symbols
represent experimental results.  In (a), continuous lines are fits
to Eq.\ (\ref{ph}), with $\epsilon =0.45 \pi$ and $N$ ranging from
$4$ for detector bias $\pm 103~\mu$V to $0$ for detector bias 0
$\mu$V.  In (b), continuous lines are fits to Eq.\ (\ref{fwhm})
(with the same $N$ and $\epsilon$) for same detector bias.}
\end{figure}


\begin{references}
\bibitem{epr} A. Einstein, B. Podolsky and N. Rosen, {\it Phys. Rev.} {\bf
47}, 777, (1935).

\bibitem{scat} E. Schr\"odinger, {\it Naturwiss.} {\bf 48}, 807, 823 and
844 (1935); {\it Proc. Am. Phil. Soc.} {\bf 124}, 323 (1980).

\bibitem{bell} J. S. Bell, {\it Physics} {\bf 1}, 195 (1964).

\bibitem{qcode} S. Weisner, {\it Sigact News} {\bf 15}, 78 (1983);  A. K.
Ekert, {\it Phys. Rev. Lett.} {\bf 67}, 661 (1991).

\bibitem{tp} C. H. Bennett et al., {\it Phys. Rev. Lett.} {\bf 70}, 1895 (1993).

\bibitem{qdc} R. Jozsa and B. Schumacher, {\it J. Mod. Optics} {\bf 41},
2343 (1994); B. Schumacher, {\it Phys. Rev.} {\bf A51}, 2738 (1995).

\bibitem{qcom} See {\it Introduction to Quantum Computation and
Information}; eds. H.-K. Lo, S. Popescu, and T. P. Spiller (Singapore:
World Scientific), 1998.

\bibitem{icqox} For a brief review see D. Rohrlich, {\it Op. Spec.} {\bf 99}, 503 (2005).

\bibitem{yacoby} A. Yacoby et al., {\it
Phys. Rev. Lett.} {\bf 74}, 4047 (1995).

\bibitem{schus} R. Schuster et al., {\it Nature} {\bf 385} 417 (1997).

\bibitem{sprinzak} D. Sprinzak et al.,
{\it Phys. Rev. Lett.} {\bf 84}, 5820 (2000).

\bibitem{ji} Y. Ji et al., {\it Nature} {\bf 422}, 415 (2003).

\bibitem{fields} M. Field et al.,
{\it Phys. Rev. Lett.} {\bf 70}, 1311 (1993).

\bibitem{buks} E. Buks et al., {\it Nature} {\bf 391}, 871 (1998).

\bibitem{feyn} R. P.  Feynman, R. B. Leighton and M. Sands, {\it The
Feynman Lectures on Physics, Vol. {\bf III}} (Reading, MA: Addison--Wesley Pub. Co.),
1965, p. 1-9.  See also Y. Aharonov and D. Rohrlich, {\it Quantum Paradoxes: Quantum
Mechanics for the Perplexed} (Weinheim:  Wiley-VCH), 2005, Sect. 18.1.

\bibitem{sai} A. Stern, Y. Aharonov and Y. Imry, {\it Phys. Rev.} {\bf
A41}, 3436 (1990).

\bibitem{noise} For a review see M. Reznikov et al., {\it Superlattices
and Microstructures} {\bf 23}, 901 (1998).  The term ``shot noise" is
often identified with the factor $NT(1-T)$.

\bibitem{eps} We rewrite $e^{i \sum_{k=j^\prime N}^{jN} \epsilon_k}$ as $\prod_{k=j^\prime
N}^{jN}e^{i \epsilon_k}$, and the expectation value of $e^{i\epsilon_k}$ is
$R+e^{i\epsilon}T$.

\bibitem{davies} J. H. Davies, J. C. Egues and J. W. Wilkins, {\it Phys.
Rev.} {\bf B52}, 11259 (1995).  For $\epsilon$ small, we have $\vert R+e^{i\epsilon}
T\vert^N \approx e^{-NRT\epsilon^2/2}=e^{-NT(1-T)\epsilon^2/2}$.

\bibitem{picc} R. de-Picciotto et al., {\it Nature} {\bf 389}, 162 (1997);
M. Reznikov et al., {\it Nature} {\bf 399}, 238 (1999).

\bibitem{lever} By applying a DC bias to the tunneling current, we shift the energies
of the electrons at resonance by a known amount.  The shift shows up as a shift in the
Coulomb blockade peaks.  Comparing this calibrated shift with the energy scale defined by
the plunger gate potential, we extract a levering factor with which we convert the width
in {\it plunger gate potential} of a Coulomb blockade peak into width in {\it electron
energy}.

\bibitem{laps} We recognize the factor $1/2 +R_1R_2/(1-R_1R_2)$ in Eq.\ (\ref{phe}) as
the average number of laps of a tunneling electron, as calculated.

\end{references}
\end{document}